\documentclass[%
 reprint,
 amsmath,amssymb,
 aps,pre,
]{revtex4-2}
\usepackage{graphicx}
\usepackage{dcolumn}
\usepackage{bm}
\usepackage{subcaption}
\usepackage{hyperref}
\usepackage{xcolor}
\usepackage{amsmath}
\usepackage{amsthm} 
\theoremstyle{plain} 
\newtheorem{thm}{Theorem}
\newtheorem{lem}{Lemma}
\theoremstyle{definition} 

\theoremstyle{remark} 
\DeclareMathOperator*{\argmin}{argmin}
\begin{document}
\preprint{APS/123-QED}
\title{Existence and Uniqueness of Nearest Stealthy Hyperuniform Configurations from Random Initial Conditions}
\author{Fausto Martelli}
 \email{fausto.martelli@cnr.it}
\affiliation{%
 CNR-Istituto dei Sistemi Complessi, P.le Aldo Moro 5, 00185 Rome (Italy)
}%
\affiliation{
 Dipartimento di Fisica, Universit\'a La Sapienza di Roma, P.le Aldo Moro 5, 00185 Rome (Italy)
}%
\date{\today}
\begin{abstract}
We prove that, for almost every uniformly random configuration of $N$ particles in a two-dimensional periodic square domain, there exists a unique nearest stealthy hyperuniform configuration, up to floppy-mode displacements, when a finite set of $m$ low-$k$ Fourier modes is constrained to vanish exactly. Under an explicit full-rank Jacobian assumption, the displacement component orthogonal to the tangent space of the finite-mode hyperuniform manifold is uniquely determined, while residual tangent-space degrees of freedom correspond to first-order floppy modes that preserve the constrained low-$k$ density fluctuations. We restrict attention to the regime of constrained-mode fraction $\chi=m/(dN)$ in which such configurations are known to remain disordered rather than crystallize. To corroborate this framework, we implement a gradient-based generator network that iteratively displaces particles to minimize a combined loss functional of hyperuniformity, short-range repulsion, and smoothness, and whose dynamics is expected to approximate the theory's minimal orthogonal projection onto the hyperuniform manifold. The network drives generic random configurations toward disordered hyperuniform representatives, progressively suppressing long-wavelength density fluctuations as measured by the structure factor and number-variance exponent\textcolor{black}{, over a reciprocal-space window that extends well beyond the explicitly constrained modes and is bounded, as we show, by the structure-factor sum rule. Local bond-orientational analysis and the absence of Bragg peaks confirm that the resulting states remain disordered}. This integrated analytical and computational approach provides a rigorous geometric foundation for understanding stealthy hyperuniformity and for generating its disordered representatives from generic random configurations.
\end{abstract}
\maketitle
\section{Introduction}
Disordered hyperuniform (DHU) systems~\cite{torquato2003local} are characterized by the suppression of long-wavelength density fluctuations, expressed by the vanishing of the structure factor $S(k)$ as $k\rightarrow 0$. These states occupy a singular position between crystalline order and complete randomness and have been identified in (out-of-)equilibrium systems. Examples are as different as jammed packings, amorphous materials, mixtures, active matter, open quantum systems, biological tissues, ecosystems, and neural networks\cite{laurent08,zach11,xie2013hyperuniformity,weijs15,hexner15,martelli2017,carollo2017,lei2019,zheng2020disordered,yanagishima2021,martelli2022steady,formanek2023molecular,cengio2024,martelli2024hyperuniform,wang2025,oppenheimer2022,ge2023,hu2025,leoni2025confinement,dam2026hyperuniformity,martelli2026disordered}, among others. Despite extensive investigation of their structural and physical properties, a fundamental geometric question remains unresolved: given a generic random configuration of particles, does there exist a well-defined nearest hyperuniform configuration, is it uniquely determined, and under what conditions is it disordered rather than ordered?
Existing constructions of DHU states largely rely on either specific interaction potentials tuned to suppress long-wavelength fluctuations, or on direct reciprocal-space optimization of a finite set of collective density variables. The latter approach, known as \emph{stealthy} hyperuniformity, imposes $\hat\rho(k;X)=0$ exactly for all wave vectors within a bounded region of reciprocal space, and constructs ground states of the associated potential $\Phi(X)=\sum_{\alpha}|\hat\rho(k_\alpha;X)|^2$ by energy minimization from random initial conditions~\cite{uche2004constraints,batten2008classical,zhang2015ground1,zhang2015ground2}. Numerical studies of this construction show that the resulting ground states are highly degenerate and disordered when the fraction of constrained modes $\chi=m/(dN)$ (with $d=2$ throughout this work) is sufficiently small, but undergo a transition toward crystalline or quasicrystalline order as $\chi$ increases beyond a dimension-dependent threshold~\cite{zhang2015ground1,zhang2015ground2}. Our finite-mode constraint map $F$ coincides exactly with this stealthy condition: $\mathcal M=F^{-1}(0)$ is the zero-energy ground-state manifold of $\Phi$. What has not, to our knowledge, been established is whether this ground-state manifold is geometrically well posed as a submanifold of configuration space, and whether a generic random configuration admits a well-defined, locally unique nearest point on it. We answer this analytically via the implicit function theorem, complementing the existing numerical evidence for the accessibility of stealthy hyperuniform states with a rigorous local existence and uniqueness result. Throughout, we restrict to values of $\chi$ within the regime where stealthy ground states are known to remain disordered, and we verify numerically \textcolor{black}{that the resulting configurations are free of Bragg peaks and carry no long-range bond-orientational order} (Sec.~\ref{sec:numerics}). We treat the finite-mode stealthy condition introduced above as a smooth constraint manifold in configuration space. For simplicity, we limit to square domains in two dimensions; constraining a finite set of low-$k$ modes is the natural finite-domain realization of the asymptotic condition $S(k)\rightarrow0$, and permits the solution set $\mathcal M$ to be treated as a smooth submanifold of the $2N$-dimensional configuration space.\newline
Under a full-rank Jacobian assumption, satisfied with probability one for uniformly random configurations and established explicitly via a lattice-based witness construction (Sec.~\ref{sec:math}), we prove that the finite-mode hyperuniform states form a smooth submanifold of codimension \textcolor{black}{equal to the number of independent real constraints, that is, twice the number $m$ of constrained modes}. For almost every random initial configuration, there exists a nearby hyperuniform configuration whose displacement component orthogonal to the manifold's tangent space is uniquely determined. The residual indeterminacy corresponds to the tangent directions that preserve the constrained low-$k$ modes to first order; we refer to these as floppy-mode displacements,\footnote{We borrow this term from the theory of constraint networks and isostaticity (Maxwell counting), to distinguish these tangent-space directions from the unrelated notion of `rattler' particles (locally caged, force-free particles) used in the jamming literature.} which we verify numerically remain non-diffusive (Sec.~\ref{sec:numerics}). \newline
The uniqueness established here is local and holds within the basin of validity of the linearized constraints; it identifies a unique hyperuniform equivalence class modulo linear floppy-mode displacements.
Our framework admits a natural interpretation within the inherent-structure paradigm introduced by Stillinger and Weber~\cite{stillinger1982hidden,stillinger1984packing}. In the potential-energy landscape framework, each configuration of a liquid (or of a glass) maps to a unique local minimum under steepest descent, defining its inherent structure. Here, hyperuniformity does not define an energy minimum but a geometric constraint manifold in the configuration space. We show that almost every random configuration admits a locally unique projection onto this manifold, thereby defining a hyperuniform inherent structure determined by minimal Euclidean displacement rather than by energetic relaxation. In this sense, hyperuniformity emerges as a geometric inherent structure of random particle arrangements.
To corroborate the analytical result, we implement a generative differentiable gradient-based scheme that minimizes a loss functional combining low-$k$ suppression of the structure factor, short-range repulsion, and smoothness regularization. The resulting dynamics performs a constructive local projection onto the hyperuniform manifold, expected to approximate the minimal orthogonal displacement predicted by the theory. Numerical experiments illustrate systematic suppression of long-wavelength density fluctuations and convergence from generic random initial conditions toward the hyperuniform manifold, within the disordered regime identified above.

By establishing the local existence and uniqueness of hyperuniform projections and connecting this structure to inherent-structure theory, this work provides a geometric foundation for understanding (disordered) hyperuniformity as a canonical organization underlying generic random configurations.
\section{\label{sec:math}Mathematical formulation}
Consider $N$ particles with positions $X=(x_1, \ldots, x_N) \in (\mathbb{T}^2)^N$, where each $x_j$ is drawn independently and uniformly on the flat torus $\mathbb{T}^2 = \mathbb{R}^2/L\mathbb{Z}^2$, $L$ being the size of the initial square domain. For a wave vector $k \in \frac{2\pi}{L}\mathbb{Z}^2 \setminus \{0\}$, define the (complex) collective density mode $\hat{\rho}(k; X) = \sum_{j=1}^N e^{-i k \cdot x_j}$, so that the structure factor is $S(k;X) = \frac{1}{N}|\hat{\rho}(k;X)|^2$.
Let $\{k_\alpha\}_{\alpha=1}^{m}$ be a collection of wave vectors with $0 < |k_\alpha| < k_c$, chosen so that no two are related by $k_\alpha = -k_\beta$ (to avoid redundancy from $\hat{\rho}(-k) = \overline{\hat{\rho}(k)}$). Define the constraint map $F: (\mathbb{T}^2)^N \to \mathbb{R}^{2m}$ by $F(X) = \bigl(\operatorname{Re}\hat{\rho}(k_1;X), \operatorname{Im}\hat{\rho}(k_1;X), \ldots, \operatorname{Re}\hat{\rho}(k_m;X), \operatorname{Im}\hat{\rho}(k_m;X)\bigr)$. The finite-mode hyperuniform manifold is $\mathcal{M} = F^{-1}(0)$, i.e., the set of configurations for which $S(k_\alpha; X) = 0$ for all $\alpha = 1, \ldots, m$.\newline
\textbf{Assumption} (Full Rank). $m < N$, and the exceptional set~\footnote{With the term \emph{exceptional} we adopt standard terminology in analysis and geometry referring to sets for which the implicit function theorem breaks down and the manifold $\mathcal{M}$ isn't guaranteed to be smooth.} $\Sigma = \{X \in (\mathbb{T}^2)^N : \operatorname{rank} DF(X) < 2m\}$ has Lebesgue measure zero.\newline
Under this assumption the implicit function theorem applies at every $X\notin\Sigma$, so the regular part $\mathcal M\setminus\Sigma$ is a real-analytic embedded submanifold of $(\mathbb T^2)^N$ of codimension $2m$, with tangent space $T_X\mathcal M=\ker DF(X)$. Since $\mathcal M$ itself has measure zero, $\Sigma$ being null does not preclude $\Sigma\cap\mathcal M\neq\emptyset$, and indeed it does not (SM, Sec.~S1). All statements below are therefore local, anchored at a random $X_0$, which lies outside $\Sigma$ almost surely; the full-rank condition being open, $DF$ retains full rank throughout a ball around such an $X_0$.

\begin{lem}[Nonemptiness and attained rank]\label{lem:witness}
Let $N=n^2$ for some $n\in\mathbb N$, let $a=L/n$, and let $X_{\rm lat}=\{(pa,qa):0\le p,q<n\}$ be the square-lattice configuration. Write $k_\alpha=\frac{2\pi}{L}(p_\alpha,q_\alpha)$ and assume the non-aliasing condition
\begin{equation}\label{eq:nonalias}
\max_\alpha\max\bigl(|p_\alpha|,|q_\alpha|\bigr)<n/2
\qquad\Longleftrightarrow\qquad
n>k_cL/\pi.
\end{equation}
Then $F(X_{\rm lat})=0$ (so $\mathcal M\neq\emptyset$), and $DF(X_{\rm lat})$ has rank $2m$; consequently $\Sigma$ is a proper real-analytic subvariety and the Assumption holds.
\end{lem}

\textcolor{black}{The proof is given in the Supplemental Material (SM), Sec.~S1.}\newline 
\textbf{Equivalence relation}. For $X, Y \in \mathcal{M}$ lying in a common tubular neighborhood, we write $X \sim Y$ if $X$ and $Y$ can be connected by a smooth path in $\mathcal{M}$ whose velocity lies in $\ker DF$ everywhere. Denote the equivalence class of $X$ by $[X]$ and the quotient by $\mathcal{M}/{\sim}$.
\begin{thm}
Let $X_0 \in (\mathbb{T}^2)^N$ be drawn from the uniform (i.i.d.) measure. Then, almost surely, the Jacobian $DF(X_0)$ has full row rank $2m$, and the following hold. \newline
(i) (Existence and local uniqueness, linearized). The minimum-norm solution of the linearized system $DF(X_0)\,\delta X = -F(X_0)$ is uniquely given by $\delta X^* = -DF(X_0)^T(DF(X_0)DF(X_0)^T)^{-1} F(X_0)$. It is the unique displacement of smallest Euclidean norm satisfying the linearized constraints at $X_0$, is orthogonal to $\ker DF(X_0)$, and any other solution differs from it by an element of $\ker DF(X_0)$. \newline
{\color{black}(ii) (Nonlinear extension: exact nearest point). There is a radius $r>0$, depending on $X_0$ only through \textcolor{black}{$\|(DF(X_0)DF(X_0)^T)^{-1}\|$ (and on the global Lipschitz constant $\mathrm{Lip}(DF)$)}, such that if $|\delta X^*|<r$ then the closed ball $B=\bar B(X_0,c|\delta X^*|)$ is disjoint from $\Sigma$ and $\mathcal M\cap B$ is a nonempty real-analytic embedded submanifold of dimension $2N-2m$ with positive reach in $B$. Consequently the metric projection
$\pi_{\mathcal M}(X_0)=\argmin\{|X-X_0| : X\in\mathcal M\cap B\}$
exists, is unique, and satisfies $X_0-\pi_{\mathcal M}(X_0)\perp T_{\pi_{\mathcal M}(X_0)}\mathcal M$: the nearest stealthy hyperuniform configuration to $X_0$ is well defined.\newline
\newline
(iii) (Constructive approximation). The Gauss--Newton (Newton--Kantorovich) iteration started at $X_0$ converges quadratically to a point $X^*\in\mathcal M\cap B$ obeying $|X^*-X_0-\delta X^*|=O(|\delta X^*|^2)$ and $|X^*-\pi_{\mathcal M}(X_0)|=O(|\delta X^*|^2)$, so that $||X^*-X_0|-\operatorname{dist}(X_0,\mathcal M)|=O(|\delta X^*|^3)$: the closed-form $\delta X^*$ is the first-order approximation of the exact projection displacement.} The tangent space $T_{X^*}\mathcal{M} = \ker DF(X^*)$ parametrizes a $(2N-2m)$-dimensional family of configurations on $\mathcal{M}$ near $X^*$ (the linear floppy-mode displacements) which are not competitors for nearest-point status but represent the residual physical degrees of freedom of the resulting hyperuniform state: the equivalence class $[X^*]\in\mathcal{M}/{\sim}$ is therefore the uniquely determined hyperuniform inherent structure associated with $X_0$.
\end{thm}
\textcolor{black}{The proof, together with the perturbation estimate identifying $\ker DF(X^*)$ with $\ker DF(X_0)$ to leading order, is given in SM, Sec.~S1.}
\begin{figure}[!h]
    \centering
    \includegraphics[width=\linewidth]{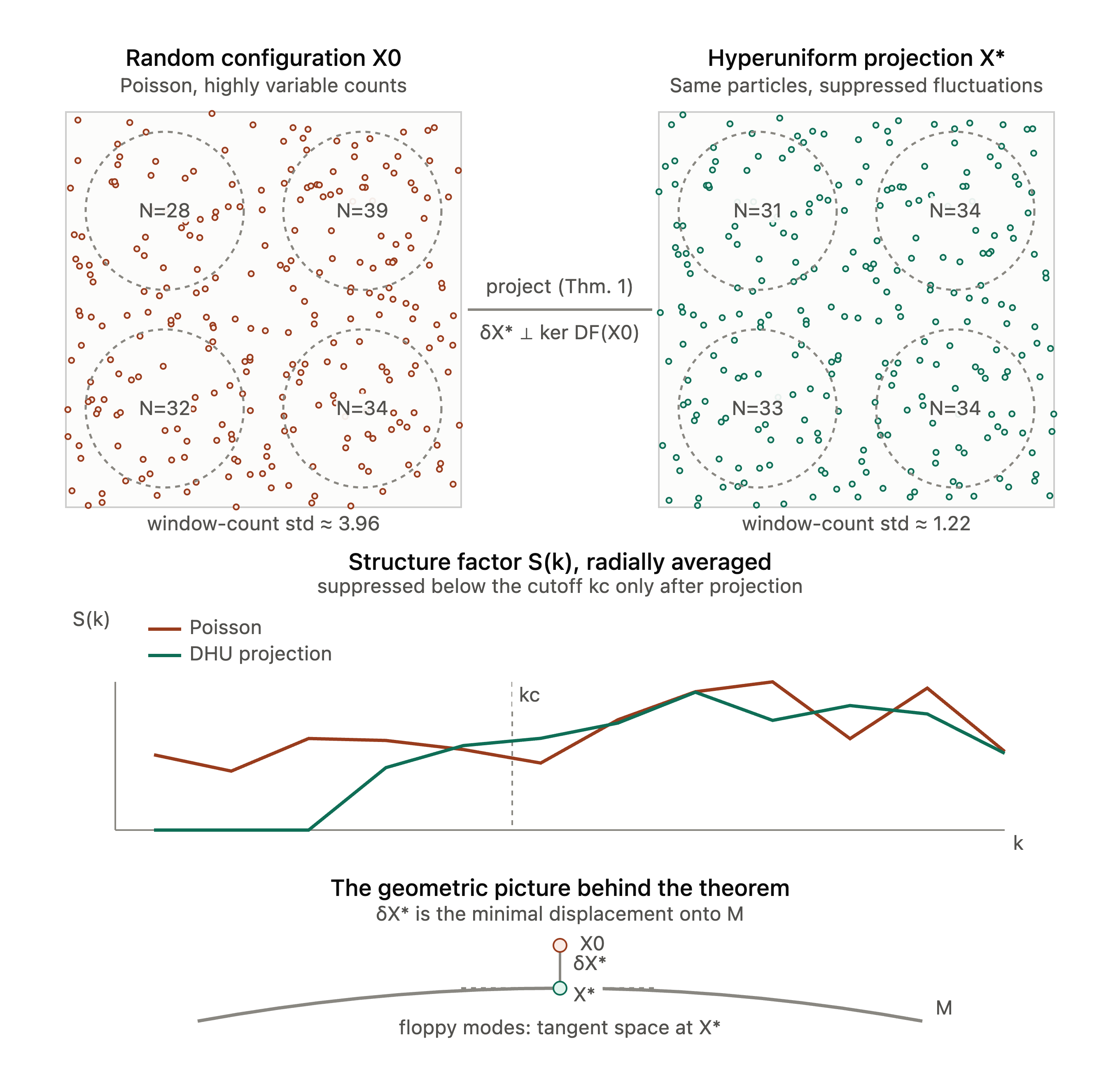}
    \caption{Closed-form illustration of Theorem~1, computed directly from the Newton--Kantorovich construction\textcolor{black}{, for a small system chosen for visual clarity}: for a configuration of $N=260$ particles on a box of side $L=20$ with $m=24$ constrained modes ($\chi\approx0.046$)\textcolor{black}{. This system is independent of, and much smaller than, the $N=10^4$ network run of Sec.~\ref{sec:numerics}}, the top row compares the random configuration $X_0$ against its hyperuniform projection $X^*$ through particle counts in four fixed sampling windows, whose window-to-window count fluctuations are visibly and quantitatively suppressed after projection. The middle panel shows the corresponding radially averaged structure factors $S(k)$, confirming suppression below the cutoff $k_c$ only for the projected configuration; the bottom panel is the abstract geometric picture underlying the theorem, in which $X_0$ is displaced by the minimal orthogonal correction $\delta X^*$ onto the nearest point $X^*$ of the hyperuniform manifold $\mathcal{M}$, with the dashed line indicating the floppy-mode tangent direction $T_{X^*}\mathcal{M}$.}
    \label{fig:scheme}
\end{figure}
Figure~\ref{fig:scheme} illustrates the geometric picture established analytically in Sec.~\ref{sec:math}. A random configuration $X_0$ is projected onto its nearest hyperuniform counterpart $X^*$; counting particles in four fixed windows shows the count fluctuations characteristic of a Poisson process in $X_0$ (window-count standard deviation $\approx3.96$) reduced markedly after projection (standard deviation $\approx1.22$)\textcolor{black}{. These two numbers are standard deviations of raw particle counts for the small illustrative system of Fig.~\ref{fig:scheme}, and are unrelated to the number-variance scaling exponent $\alpha$ reported for the $N=10^4$ network run in Sec.~\ref{sec:numerics}}, and the radially averaged structure factor $S(k)$ is correspondingly suppressed below the imposed cutoff $k_c$ only for $X^*$, while both configurations retain comparable structure at higher $k$. The bottom panel abstracts this same displacement as $X_0\mapsto X^*=X_0+\delta X^*$, with the tangent direction at $X^*$ representing the floppy-mode displacements of Theorem~1. A direct quantitative comparison between the network's converged displacement and this closed-form $\delta X^*$ is left to future work.

\section{\label{sec:numerics}Numerical experiments}
Starting from a uniformly random configuration of $N$ particles in a two-dimensional periodic box, the code generates DHU configurations by applying small, smooth particle displacements via a generator network. \textcolor{black}{All results below are for $N=10^4$ particles in a box of side $L=10^3$, so that the mean interparticle spacing is $\ell=L/\sqrt N=10$. The network architecture and training protocol are summarized in SM, Sec.~S2 and Fig.~S1.} At each iteration, the loss function $\mathcal{L}$ combines three contributions:
\begin{enumerate}
    \item Hyperuniformity term $\mathcal{L}_{hyper}=<(S(k)-S_{target})^2>_{|k|<k_c}$, which penalizes deviations of the structure factor $S(k)$ from a small target value at low wave-vectors. \textcolor{black}{The average runs over the discrete reciprocal-lattice vectors $k\in\frac{2\pi}{L}\mathbb Z^2$ with $0<|k|<k_c$, with cutoff $k_c=8\pi/L\simeq2.5\times10^{-2}$; this selects $m=22$ independent constrained modes (44 counting $\pm k$ pairs, redundant because $S(-k)=S(k)$), i.e.\ a nominal constrained-mode fraction $\chi=m/(dN)=1.1\times10^{-3}$. The target is a small positive constant, $S_{target}=10^{-4}$, rather than exactly zero: the scheme enforces the stealthy condition of Sec.~\ref{sec:math} in a soft sense, driving the configuration into a thin neighborhood of $\mathcal M$ rather than exactly onto it.}
    \item Local repulsion term $\mathcal{L}_{rep}=<1/r_{ij}^2>_{r_{ij}<r_c}$, which prevents particle overlaps and maintains short-range structural regularity\textcolor{black}{; the average is taken over all distinct pairs closer than the cutoff $r_c=25=2.5\,\ell$, in the minimum-image convention}.
    \item Smoothness term $\mathcal{L}_{smooth}=<(S(\textcolor{black}{k})-\bar{S})^2>_k$, which regularizes high-frequency fluctuations. \textcolor{black}{Here $\bar S=\langle S(k)\rangle_k$ is the mean of $S(k)$ over the entire sampled reciprocal-space grid and the outer average runs over that same grid, a square of half-width $k_{\rm grid}=60\times2\pi/L\simeq0.377$. By penalizing the variance of $S(k)$ across reciprocal space, this term suppresses the growth of sharp diffraction peaks and acts as an explicit regularizer against crystallization.}
\end{enumerate}
The total loss $\mathcal{L}=\mathcal{L}_{hyper}+\mathcal{L}_{rep}+\mathcal{L}_{smooth}$
is minimized using stochastic gradient-based optimization, which drives the initial configuration toward $\mathcal{M}$. Because the network follows the gradient of $\mathcal{L}$ rather than the exact Gauss--Newton flow underlying Theorem~1, its trajectory is expected to approximate the theory's minimal orthogonal projection onto $\mathcal{M}$; residual degrees of freedom within the tangent space correspond to small floppy-mode motions that preserve hyperuniformity to first order. A quantitative comparison between the network's converged displacement and $\delta X^*$ is left to future work.

{\color{black}We emphasize that the variable $t$ denotes the optimization epoch index, i.e., the number of completed gradient-descent updates of the network parameters, and not a physical time. The trajectory $X(t)$ is a path in configuration space generated by minimization of $\mathcal L$, with no equation of motion, no thermostat and no dynamical time step behind it. It should therefore be read as a constructive route to the hyperuniform state, not as a model of any relaxation process. We note also that the parametrization caps each Cartesian component of the per-particle displacement at $5\ell$ (Euclidean norm at $5\sqrt2\,\ell$) (SM, Sec.~S2), well outside the small-$|\delta X^*|$ regime of Theorem~1; the measured displacements, however, remain far below that cap, as the positional correlations below confirm.}

\begin{figure*}[t]
\centering
\includegraphics[width=\textwidth]{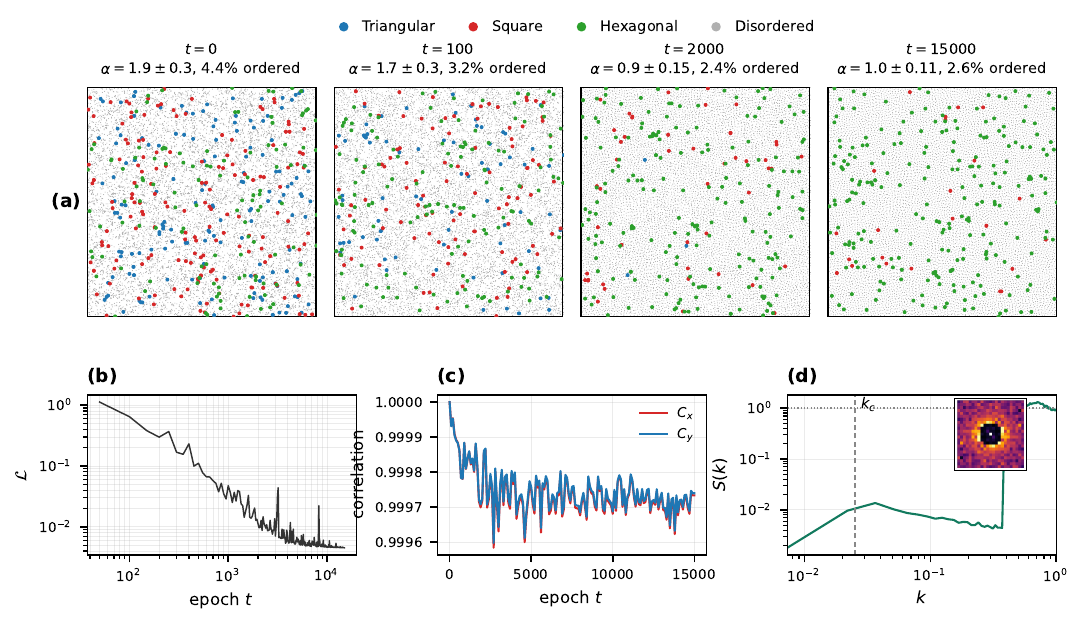}
\caption{\textcolor{black}{(a) Configuration snapshots at selected epochs $t$, with each particle colored by its local bond-orientational environment (triangular, square, hexagonal or disordered; see text). The number-variance exponents $\alpha$ are fitted over $0.06L<R<0.20L$. (b) Loss function $\mathcal L$ versus epoch on logarithmic axes. (c) Positional correlations $C_x(t)$ and $C_y(t)$ showing that particles do not diffuse away from their initial positions. (d) Angularly averaged structure factor $S(k)$ of the final configuration, with the two-dimensional $S(\mathbf k)$ map inset; the dotted line marks the ideal-gas value $S=1$ and the dashed line the cutoff $k_c$.}}
\label{fig:combined}
\end{figure*}
Figure~\ref{fig:combined}\textcolor{black}{(b)} reports the profile of the loss function during training\textcolor{black}{, on logarithmic axes}, along with representative snapshots \textcolor{black}{in Fig.~\ref{fig:combined}(a)}. The initial distribution of $N=10000$ points is randomly generated as shown by the exponent $\alpha$ of the number variance
\textcolor{black}{$\sigma^{2}(R)= \langle N(R) \rangle [ \frac{1}{(2\pi)^{d}} \int_{\mathbb{R}^d}S(k)\tilde{\omega}(k;R)dk ]$,
where $\langle N(R)\rangle=\rho\,v_1(R)$ is the mean count in a circular window of radius $R$, $v_1(R)$ is the window volume, and $\tilde\omega(k;R)$ is the Fourier transform of the scaled intersection volume of two such windows~\cite{torquato2003local}. Its large-$R$ scaling defines the exponent,
\begin{equation}
  \sigma^{2}(R)\sim R^{\alpha},
  \label{eq:alphadef}
\end{equation}
with $\alpha=d=2$ for a Poisson point pattern and $\alpha=d-1=1$ for a class-I (strongly) hyperuniform system in $d=2$, the latter being a rigorous lower bound for any point pattern.}
The exponent $\alpha$ decreases to values typical of DHU systems\textcolor{black}{, falling from $\alpha=1.9\pm0.3$ at $t=0$ to $\alpha=1.0\pm0.11$ at $t=1.5\times10^4$, i.e.\ from the Poisson value to the class-I value within the quoted uncertainty}.

{\color{black}To confirm that the final state is disordered rather than crystalline, we first classify the neighborhood of each particle in Fig.~\ref{fig:combined}(a) by its local bond-orientational order parameter~\cite{NelsonHalperin1979} $\psi_n(i)=\frac{1}{N_i}\sum_{j\in\partial i}e^{\,in\theta_{ij}}$, where the sum runs over the $N_i=6$ nearest neighbors of particle $i$ and $\theta_{ij}$ is the angle of the bond $ij$ with respect to a fixed axis. \newline 
Next, we inspect the structure factor of the final configuration. As shown in Fig.~\ref{fig:combined}(d), no Bragg peaks appear anywhere in the accessible range. We also observe a feature worth stating explicitly: suppression is not confined to $|k|<k_c$. $S(k)$ is reduced by about two orders of magnitude over the whole grid of half-width $k_{\rm grid}\simeq0.377$ sampled by $\mathcal L_{smooth}$, and recovers to unity immediately above it. The effective number of suppressed independent modes is therefore not $m=22$ but $\simeq5.7\times10^3$, corresponding to $\chi_{\rm eff}\simeq0.28$ rather than the nominal $\chi=1.1\times10^{-3}$. This remains below the $\chi\simeq0.5$ threshold above which stealthy ground states in $d=2$ order~\cite{zhang2015ground1,zhang2015ground2}, consistent with the absence of Bragg peaks and with the bond-orientational statistics above, but it places the configurations considerably closer to that transition than the nominal $\chi$ suggests. Moreover, the width of the suppressed window is bounded by the sum rule $(2\pi)^{-d}\int[S(k)-1]\,d^dk=-\rho$: a fully suppressed disk of radius $k_{\rm grid}$ would already exhaust that budget, so the observed window is essentially as wide as a point pattern of this density permits, and the compensation appears as $S(k)$ slightly exceeding unity above the grid edge.} \newline 
Particles don't diffuse during the generation of the DHU distribution, as confirmed by the Pearson correlation coefficient \textcolor{black}{between the initial and the current coordinates}, $C_{\textcolor{black}{\beta}}(t)=\frac{\text{Cov}(\textcolor{black}{\beta}(0),\textcolor{black}{\beta}(t))}{\sqrt{\text{Var}(\textcolor{black}{\beta}(0))\,\text{Var}(\textcolor{black}{\beta}(t))}}$, $\textcolor{black}{\beta}=x,y$,
where $t$ is the progression of epochs. Figure~\ref{fig:combined}\textcolor{black}{(c)} shows that both coefficients deviate only marginally from $1$. \textcolor{black}{The two directions track one another closely, and the residual difference between them changes sign when the network is retrained with a different random seed, identifying it as a single-realization fluctuation rather than an anisotropy introduced by the algorithm. This is consistent with the loss being isotropic: $\mathcal L_{hyper}$ and $\mathcal L_{smooth}$ are averages over a symmetric grid and $\mathcal L_{rep}$ depends only on interparticle distances, while nothing in the generator parametrization enforces exact statistical isotropy at finite $N$.}

\section{Conclusions}
We have established a rigorous mathematical foundation for stealthy hyperuniform configurations constructed from random initial conditions. By formulating hyperuniformity as a finite set of smooth Fourier-space constraints (equivalent to the zero-energy ground-state condition of $\Phi(X)=\sum_\alpha|\hat\rho(k_\alpha;X)|^2$ studied numerically in Refs.~\cite{uche2004constraints,batten2008classical,zhang2015ground1,zhang2015ground2}) we showed that the corresponding solution set forms\textcolor{black}{, away from an exceptional set of measure zero,} a smooth submanifold whose tangent space encodes all first-order floppy-mode motions, and we established explicitly, via a lattice witness construction (Lemma~\ref{lem:witness}), that (i) this manifold is nonempty, and (ii) the Jacobian rank required by the theorem is attained. For almost every random initial configuration, \textcolor{black}{the nearest point of this manifold exists and is unique}, yielding a well-defined minimal-displacement hyperuniform inherent structure, unique up to the equivalence class $[X^*]\in\mathcal{M}/{\sim}$\textcolor{black}{, and computable to leading order in closed form}. This result clarifies the geometric nature of stealthy hyperuniformity and provides a principled method for constructing hyperuniform states, disordered within the regime characterized by $\chi$, from arbitrary initial conditions.
Our numerical experiments are consistent with this framework: gradient-based optimization drives generic random configurations toward the hyperuniform manifold, while the evolution of the number variance exponent confirms the progressive suppression of long-wavelength fluctuations\textcolor{black}{, and both the local bond-orientational statistics and the absence of Bragg peaks confirm that the resulting states remain disordered}.
This work opens several avenues for future research, including a quantitative comparison between the network's converged displacement and the closed-form projection of Theorem~1, and the characterization of the global geometry of the hyperuniform manifold and its dependence on $\chi$, in particular the relation between the local uniqueness established here and the known order-disorder transition at larger $\chi$\textcolor{black}{, which the effective value $\chi_{\rm eff}\simeq0.28$ reached here brings within reach}. The role of higher-order constraints, and the extension of the inherent-structure concept to dynamical or thermodynamic settings, are also avenues for future research. More broadly, the framework presented here provides a bridge between hyperuniformity, optimization, and geometric analysis.

\textcolor{black}{\emph{Data availability.} The code implementing the generator scheme is openly available at \url{https://github.com/fausto-martelli/Disordered-Hyperuniformity-Generator}.}

\bibliography{apssamp}

\clearpage
\onecolumngrid
\setcounter{page}{1}
\setcounter{equation}{0}
\setcounter{figure}{0}
\setcounter{section}{0}
\renewcommand{\theequation}{S\arabic{equation}}
\renewcommand{\thefigure}{S\arabic{figure}}
\renewcommand{\thesection}{S\arabic{section}}
\begin{center}
{\large\textbf{Supplemental Material}}\\[2pt]
{\large Existence and Uniqueness of Nearest Stealthy Hyperuniform Configurations\\ from Random Initial Conditions}\\[4pt]
Fausto Martelli
\end{center}
\vspace{6pt}

\section{\label{sm:proofs}Proofs}

\subsection{Singular points of $\mathcal M$}
The Assumption of the main text states that $\Sigma=\{X:\operatorname{rank}DF(X)<2m\}$ is Lebesgue-null in $(\mathbb T^2)^N$. Because $\mathcal M$ is itself a null set, this does not imply $\Sigma\cap\mathcal M=\emptyset$. A single constrained mode suffices to see that the intersection is generally nonempty: write $\theta_j=k_1\cdot x_j$ and suppose every $\theta_j$ takes one of two values differing by $\pi$, with the particles split evenly between them. Then $\hat\rho(k_1)=0$, so $X\in\mathcal M$, while the two rows of $DF$ are $\mathbf s\otimes k_1$ and $\mathbf c\otimes k_1$ with $\mathbf s,\mathbf c$ proportional, so $\operatorname{rank}DF(X)=1<2m=2$. Such configurations form a positive-dimensional family. Consequently $\mathcal M$ is a manifold only away from $\Sigma$, and the results of the main text are local statements anchored at a random $X_0\notin\Sigma$.

\subsection{Proof of Lemma 1}
Since $0<\max(|p_\alpha|,|q_\alpha|)<n/2<n$ and $(p_\alpha,q_\alpha)\neq(0,0)$, at least one index is not a multiple of $n$. As
$\hat\rho(k_\alpha;X_{\rm lat})=\bigl(\sum_{p=0}^{n-1}e^{-2\pi i p_\alpha p/n}\bigr)\bigl(\sum_{q=0}^{n-1}e^{-2\pi i q_\alpha q/n}\bigr)$
and each geometric sum vanishes unless $n$ divides the corresponding index, at least one factor is zero for every $\alpha$, giving $F(X_{\rm lat})=0$.

For the rank, let $\chi_\alpha(j)=e^{ik_\alpha\cdot x_j}$ be the restriction of the plane wave to the lattice sites. These are characters of the finite abelian group $(\mathbb Z_n)^2$, and two characters coincide precisely when their index pairs agree modulo $n$. The non-aliasing condition bounds every index by $n/2$ in absolute value, so $p_\alpha\pm p_\beta$ and $q_\alpha\pm q_\beta$ cannot both vanish modulo $n$ for $\alpha\neq\beta$, and $2k_\alpha\not\equiv0$; hence the $2m$ characters $\{\chi_\alpha,\overline{\chi_\alpha}\}_{\alpha=1}^m$ are pairwise distinct, therefore orthogonal, therefore linearly independent as functions on the lattice. The rows of $DF(X_{\rm lat})$ are $-\mathbf s^\alpha\otimes k_\alpha$ and $-\mathbf c^\alpha\otimes k_\alpha$, with $\mathbf s^\alpha_j=\sin(k_\alpha\cdot x_j)$ and $\mathbf c^\alpha_j=\cos(k_\alpha\cdot x_j)$. A vanishing linear combination $\sum_\alpha(a_\alpha\mathbf s^\alpha+b_\alpha\mathbf c^\alpha)\otimes k_\alpha=0$ reads, site by site,
$\sum_\alpha\frac12[(b_\alpha-ia_\alpha)\chi_\alpha(j)+(b_\alpha+ia_\alpha)\overline{\chi_\alpha}(j)]k_\alpha=0$;
linear independence of the characters forces $(b_\alpha-ia_\alpha)k_\alpha=0$ for each $\alpha$, and $k_\alpha\neq0$ gives $a_\alpha=b_\alpha=0$. The $2m$ rows are thus linearly independent and $\operatorname{rank}DF(X_{\rm lat})=2m$.

Finally, $X\mapsto\sum(\text{squared }2m\times2m\text{ minors of }DF(X))$ is real-analytic on the connected manifold $(\mathbb T^2)^N$ and nonzero at $X_{\rm lat}$; hence it cannot vanish identically, its zero set $\Sigma$ is a proper real-analytic subvariety, and $\Sigma$ has Lebesgue measure zero. $\blacksquare$

The non-aliasing condition is necessary, not merely convenient. For $n=6$, $L=10$ and $k_c=2$ one has $m=18$ but $\operatorname{rank}DF(X_{\rm lat})=34<36=2m$, the deficit arising from the self-aliased mode $(3,0)$ and the aliased pair $(3,1),(3,-1)$.

\subsection{Proof of Theorem 1}
\emph{Part (i).} Since $F$ is real-analytic, the set where $DF$ drops rank is a proper analytic subvariety of $(\mathbb{T}^2)^N$ (Lemma 1), hence Lebesgue-null, and $DF(X_0)$ has full row rank almost surely. Given full rank, the linear system $DF(X_0)\delta X=-F(X_0)$ is consistent and its general solution is $\delta X=\delta X^*+v$ with $v\in\ker DF(X_0)$. By the normal equations $\delta X^*$ is the unique element of minimal norm, and $\delta X^*\perp\ker DF(X_0)$.

\emph{Part (ii).} Because rank is lower semicontinuous, the set of full-rank configurations is open; hence $DF$ retains full rank on a ball $B$ about $X_0$ whose radius is bounded below in terms of $\sigma_{\min}(DF(X_0))$ and $\mathrm{Lip}(DF)$, and $B\cap\Sigma=\emptyset$. Measure zero of $\Sigma$ alone would not suffice here: it gives density of full-rank points, not persistence. On $B$, $0$ is a regular value of $F$, so $\mathcal M\cap B$ is a real-analytic embedded submanifold of dimension $2N-2m$; it is nonempty for $|\delta X^*|$ small by the Newton--Kantorovich theorem applied at $X_0+\delta X^*$, using $F(X_0+\delta X^*)=O(|\delta X^*|^2)$. A $C^2$ embedded submanifold has positive reach locally, so the tubular neighborhood theorem yields a unique metric projection for every point of $B$ sufficiently close to $\mathcal M$, characterized by the stated orthogonality condition; $\operatorname{dist}(X_0,\mathcal M)\le|\delta X^*|+O(|\delta X^*|^2)$ places $X_0$ in that regime.

\emph{Part (iii).} Quadratic convergence and $|X^*-X_0-\delta X^*|=O(|\delta X^*|^2)$ are the standard Newton--Kantorovich estimates. The Gauss--Newton limit and the metric projection agree only to leading order: each Gauss--Newton step lies in the row space of $DF$ evaluated at the current iterate, and these row spaces differ by $O(|\delta X^*|)$ rotations, so the accumulated displacement acquires a tangential component at $X^*$ of size $O(|\delta X^*|^2)$, whence $|X^*-\pi_{\mathcal M}(X_0)|=O(|\delta X^*|^2)$. The distance from $X_0$ is stationary at the minimum, so a tangential offset $\delta$ from $\pi_{\mathcal M}(X_0)$ costs an excess $\simeq\delta^2/2\operatorname{dist}(X_0,\mathcal M)$; with $\delta=O(|\delta X^*|^2)$ and $\operatorname{dist}=O(|\delta X^*|)$ the excess is $O(|\delta X^*|^3)$. Direct numerical evaluation of the two constructions confirms both scalings over two decades in $|\delta X^*|$ (four in the tangential component). $\blacksquare$

\subsection{Identification of the tangent spaces}
Since $F$ is real-analytic on the compact manifold $(\mathbb{T}^2)^N$, $DF$ is Lipschitz with some constant $\mathrm{Lip}(DF)<\infty$, so $\|DF(X^*)-DF(X_0)\|\le\mathrm{Lip}(DF)|X^*-X_0|=O(|\delta X^*|)$. By standard perturbation bounds for the kernel of a full-rank matrix, $\ker DF(X^*)$ and $\ker DF(X_0)$ agree up to a rotation of order $O(|\delta X^*|)$, justifying the identification of floppy-mode directions at $X_0$ and at $X^*$ to the order of approximation used throughout. Note also that the two global translations always belong to $\ker DF$ at points of $\mathcal M$, since a rigid shift multiplies every $\hat\rho(k_\alpha)$ by a phase; of the $2N-2m$ tangent directions, $2N-2m-2$ are therefore nontrivial floppy modes.

\section{\label{sm:code}Numerical implementation}

\begin{figure}[h]
\centering
\includegraphics[width=0.92\textwidth]{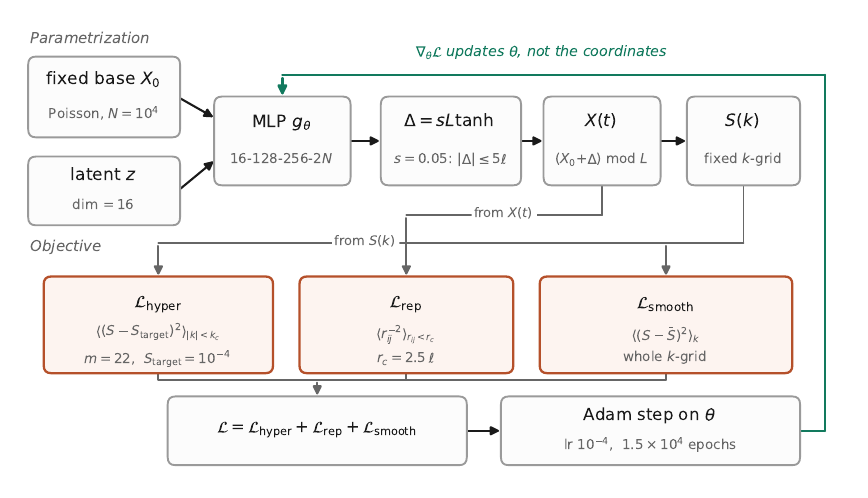}
\caption{Workflow of the generator scheme. A fixed Poisson configuration $X_0$ is never trained; the network $g_\theta$ maps a latent vector $z$ to a bounded displacement field $\Delta=sL\tanh(\cdot)$, and the particle coordinates $X(t)=(X_0+\Delta)\bmod L$ are a deterministic function of the network parameters $\theta$. The structure factor is evaluated on a fixed reciprocal-space grid and enters two of the three loss terms; the third acts in real space. The gradient of the total loss updates $\theta$, not the coordinates, so the scheme is a reparametrized optimization over displacement fields rather than a model of a distribution.}
\label{fig:workflow}
\end{figure}

The generator is a multilayer perceptron $g_\theta$ with layer widths $16\to128\to256\to2N$ and ReLU activations, mapping a latent vector $z\in\mathbb R^{16}$ to a displacement field. Base positions $X_0$ are drawn once, uniformly in the box, and stored as a fixed buffer: they are not optimized. The network output is passed through a hyperbolic tangent and rescaled,
\begin{equation}
X(t)=\bigl(X_0+sL\tanh[g_\theta(z)]\bigr)\bmod L,\qquad s=0.05,
\label{eq:param}
\end{equation}
so that \textcolor{black}{each Cartesian component of the per-particle displacement is bounded by $sL=5\ell$ (the Euclidean norm by $5\sqrt2\,\ell$)} and the configuration remains on the torus by construction. \textcolor{black}{An isotropic Gaussian perturbation of amplitude $5\times10^{-3}$, i.e.\ $5\times10^{-4}\,\ell$, is added to the coordinates at every epoch; it is negligible on every scale considered here and serves only to break exact degeneracies.} Equation~\eqref{eq:param} makes the coordinates a deterministic function of $\theta$ \textcolor{black}{up to this perturbation}; the optimization is therefore a reparametrized search over displacement fields, and the gradient updates $\theta$ rather than the particles directly (Fig.~\ref{fig:workflow}).

The structure factor is evaluated at each step on a fixed grid of $120\times120$ reciprocal-lattice vectors, i.e.\ a square of half-width $k_{\rm grid}=60\times2\pi/L$. The three loss contributions defined in the main text are summed with unit weights and minimized with Adam at learning rate $10^{-4}$ for $1.5\times10^{4}$ epochs, with gradient-norm clipping. A single latent sample is drawn per epoch. The bound $sL=5\ell$ in Eq.~\eqref{eq:param} is a cap, not a typical value: \textcolor{black}{the root-mean-square minimum-image displacement of the final configuration is [RMS]$\,\ell$, with a maximum of [MAX]$\,\ell$ over all particles, an order of magnitude below the cap}, which is what places the trajectory within the local regime assumed by Theorem~1.

\end{document}